\documentclass[aps,prd,preprintnumbers,showpacs,superscriptaddress,nofootinbib,amsmath,amssymb,floats,floatfix,showkeys,notitlepage,longbibliography]{revtex4-1}
\usepackage{comment}
\usepackage{graphicx}
\usepackage{subfigure}
\usepackage{palatino}
\usepackage[commandnameprefix=always]{changes}
\usepackage{hyperref}
\hypersetup{colorlinks=true,linkcolor=red,urlcolor=red,citecolor=red}
\usepackage[toc,page]{appendix}
\usepackage[normalem]{ulem}

\usepackage{orcidlink}
\usepackage{lipsum}
\usepackage{graphicx}
\usepackage{subfigure}
\usepackage{palatino}
\usepackage{float}
\usepackage{sans}
\usepackage{adjustbox}
\usepackage{latexsym}
\usepackage{amsmath}
\usepackage{amssymb}
\usepackage{amsfonts}
\usepackage{dcolumn}
\usepackage{bm}
\usepackage{tikz}
\usepackage{bigints}
\usepackage{array,tabularx,multirow,booktabs}
\usepackage[tracking=true]{microtype}
\SetTracking{}{500}
\SetTracking{encoding={*}, shape=sc}{40}
\UseRawInputEncoding %for inputenc error%
\allowdisplaybreaks
\usepackage{adjustbox}
\usepackage{latexsym}
\usepackage{amsmath}
\usepackage{amssymb}
\usepackage{amsfonts}
\usepackage{dcolumn}
\usepackage{bm}
\usepackage{tikz}
\usepackage{bigints}
\usepackage{array,tabularx,multirow,booktabs}
\usepackage[tracking=true]{microtype}
\usepackage{color}
\UseRawInputEncoding %for inputenc error%
\allowdisplaybreaks

\begin{document}

\title{Thermodynamic Topology, Photon Spheres, and Evidence for Weak Gravity Conjecture in Charged Black Holes with Perfect Fluid within Rastall Theory}

\author{Saeed Noori Gashti
\orcidlink{0000-0001-7844-2640}
}
\email{saeed.noorigashti@stu.umz.ac.ir}
\affiliation{School of Physics, Damghan University,\\ P. O. Box 3671641167, Damghan, Iran}

\author{\.{I}zzet Sakall{\i}
\orcidlink{0000-0001-7827-9476}
}
\email{izzet.sakallı@emu.edu.tr}
\affiliation{Physics Department, Eastern Mediterranean
University, Famagusta, 99628 North Cyprus, via Mersin 10, Turkiye}

\author{Behnam Pourhassan
\orcidlink{0000-0003-1338-7083}}
\email{b.pourhassan@du.ac.ir}
\affiliation{School of Physics, Damghan University, Damghan 3671645667, Iran.}
\affiliation{Center for Theoretical Physics, Khazar University, 41 Mehseti Street, Baku, AZ1096, Azerbaijan}
\affiliation{Physics Department, Istanbul Technical University,
Istanbul 34469, Turkey}
\affiliation{Centre of Research Impact and Outcome, Chitkara University, Rajpura-140401, Punjab, India.}

\begin{abstract}
In this paper, we explore the Weak Gravity Conjecture (WGC) within the context of photon spheres in charged black holes, framed by Perfect Fluid in Rastall Theory. We aim to validate the WGC by identifying the extremality states of these black holes.
We highlight the interplay between quantum dynamics and gravitational forces, opening new avenues in high-energy physics and quantum gravity. Our analysis reveals significant system changes with varying perfect fluid intensity (\(\alpha\)) and Rastall parameters (\(k\) and \(\lambda\)). For dust field (\(\omega = 0\)), the WGC is met in extremality (\(T=0\)) with \((Q/M)_{ext}>1\), indicating a black hole due to the presence of photon spheres (PS) with a total charge$-1$. However, further increases in \(k\) and \(\lambda\) or decreases (\(\alpha\)) lead to $PS=0$ that determines a singularity, not a black hole.
We observed that the radiation field(\(\omega = 1/3\)), quintessence field (\(\omega = -2/3\)), and phantom fields field (\(\omega = -4/3\)) also confirmed the WGC and maintaining a total charge of \(PS = -1\) in some regions of the free parameters. Our numerical solutions identify points satisfying the WGC, establishing a bridge between quantum and cosmic realms. The results are summarized in Table (\ref{P1}). We also examine Duan’s topological current \(\phi\)-mapping theory by analyzing generalized Helmholtz free energy methods to study the topological classes of our black hole. We reveal that for given values of the free parameters, the total topological numbers (\(W = 0\)) exist for the generalized Helmholtz free energy method for \(\omega = 0, 1/3, -2/3, -4/3\).
\end{abstract}

\date{\today}

\keywords{Photon Spheres, Charged Black Holes, Rastall Theory, Weak Gravity Conjecture.}

\pacs{}

\maketitle
%\tableofcontents

\section{Introduction}
The Swampland Program has emerged as a compelling framework within theoretical physics, aimed at discerning which low-energy effective field theories can be consistently embedded in a theory of quantum gravity, such as string theory, and which cannot. The theories that fail to satisfy these criteria are said to reside in the so-called "swampland," a conceptual space where inconsistent theories lie. This separation is crucial because it provides theoretical physicists with guidelines to avoid developing models that cannot be reconciled with the underlying principles of quantum gravity. At the heart of this boundary lies a collection of conjectures, which are integral to the Swampland Program, including the well-known WGC. These conjectures offer significant insights into various domains of theoretical physics, particularly in black hole physics, particle physics, and cosmology \cite{a',b',c',d',e',f',isSadeghi:2020ciy,isSadeghi:2023cxh}.

The WGC posits that, in any consistent theory of quantum gravity, gravity must be the weakest of all forces. This seemingly simple statement has profound implications. For instance, it suggests that for charged particles, the charge-to-mass ratio must exceed that of an extremal black hole. In other words, extremal black holes, which are theoretical constructs with the maximum possible charge or angular momentum for a given mass, must possess a charge-to-mass ratio equal to or more than unity, allowing other forces to dominate over gravitational interactions in certain contexts. Consequently, the WGC has far-reaching implications for our understanding of the fundamental structure of the universe, from the scale of elementary particles to the cosmological realm \cite{isHawking:1991nk,isCheung:2014vva}.

The practical applications of the WGC extend well beyond its theoretical origin. In particle physics, the WGC imposes constraints on the mass and charge of particles, suggesting that particles with large charge-to-mass ratios should exist in nature. In cosmology, the WGC can inform models of inflation and early universe dynamics by setting upper limits on field strengths and potential gradients. Black hole thermodynamics, quantum gravity, and the search for a theory beyond the Standard Model also benefit from the constraints provided by the WGC. Indeed, the WGC has been invoked in numerous studies to guide the development of consistent theories that bridge quantum mechanics, gravity, and high-energy physics, offering theoretical potential resolutions to some problems and has a wide range of studies in black hole physics, particle physics, cosmology, thermodynamics, etc\cite{a,b,f,h,i,k,l,o,p,s,v,x,z,bb,cc,dd,ee,ff,hh,ii,jj,ll,mm,nn,rr,ss,tt,uu,ww,aaa,bbb,ccc,ddd,eee,hhh,jjj,kkk,lll,mmm,ooo,ppp,qqq}.

Recently, novel methods have been developed to analyze and compute critical points and phase transitions in black hole thermodynamics. These advancements include the introduction of topological approaches, which allow for a more detailed understanding of black hole dynamics, phase transitions, and stability \cite{18a,19a,20a,21a,22a,23,24,25,26,27,28,29,31,33,34,35,37,38,39,40,41,42,43,44,44c,44d}. In particular, the topological study of photon spheres in black holes has proven to be a rich and fascinating avenue of research. The existence of photon spheres, or unstable circular orbits for light, is an essential feature of ultra-compact gravitational objects such as black holes. The study of photon spheres provides valuable insights into the structure of spacetime near black holes and is closely tied to fundamental concepts such as the Weak Cosmic Censorship Conjecture (WCCC) and gravitational lensing \cite{7,8,9,10,11,12,13,14,15,16,17,18,19,20,21}.

Photon spheres are particularly significant because they influence the shadow of a black hole, which is the observable dark region surrounding the event horizon, formed by the gravitational bending of light. The presence and stability of these photon spheres are indicative of the underlying spacetime structure and have been linked to the stability of black holes themselves. Probing these spheres can thus provide evidence for or against the WGC. If a black hole possesses photon spheres that satisfy the WGC, this suggests that the black hole conforms to the principles of quantum gravity, whereas the absence of such spheres or deviations from the WGC might indicate new physics beyond the standard model \cite{22,22',22''}.

The model of perfect fluid dark matter (PFDM) offers another interesting avenue for studying black holes. In this model, black holes are surrounded by perfect fluid matter, which can significantly alter their properties, particularly in the context of modified theories of gravity such as Rastall gravity \cite{21b,22b,23b}. Rastall gravity modifies the standard conservation laws of general relativity, allowing for a non-conserved energy-momentum tensor, which could have profound implications for cosmology and black hole physics \cite{27b,28b,29b,30b,31b,38b}. This modification introduces additional parameters, such as \(k\) and \(\lambda\), that control the degree of deviation from standard Einstein gravity, providing a broader framework within which to test the WGC and other fundamental conjectures.

Our study aims to explore the WGC in the context of charged black holes surrounded by perfect fluid dark matter, as described by Rastall gravity. Specifically, we aim to examine the extremality conditions of these black holes and determine whether they satisfy the WGC. Our results show that varying the intensity of the perfect fluid (\(\alpha\)) and the Rastall parameters (\(k\) and \(\lambda\)) leads to significant changes in the system's behavior. In particular, we find that the WGC is satisfied in extremal black holes with \((Q/M)_{ext}>1\) under certain conditions, but increasing the Rastall parameters can lead to the disappearance of photon spheres, signaling the transition from a black hole to a naked singularity.

The paper is structured as follows: In Section 2, we introduce the charged black hole solutions in Rastall gravity with perfect fluid matter. In Section 3, we investigate the photon sphere structure of these black holes and derive the conditions under which the WGC and the WCCC are satisfied. We present a detailed analysis of our results and discuss the implications for black hole thermodynamics and quantum gravity. Finally, in Section 4, we summarize our findings and suggest directions for future research.

\section{Charged Black Holes in Perfect Fluid within Rastall Theory}
In this section, we explore the solutions for charged black holes in the presence of perfect fluid matter within the framework of Rastall gravity. Rastall gravity modifies the traditional conservation law for the energy-momentum tensor, departing from the standard form in Einstein's General Relativity (GR). In this modified theory, the energy-momentum conservation is altered to the following form \cite{27b,28b}:
\begin{equation}\label{1}
T^{\mu\nu}_{;\mu} = \lambda R^{,\nu},
\end{equation}
where \(T^{\mu\nu}\) is the energy-momentum tensor, \(R\) is the Ricci scalar, and \(\lambda\) is the Rastall parameter, which reflects the extent to which the conservation of energy and momentum is modified. This modification is motivated by the idea that the geometry of spacetime could influence the energy distribution in strong gravitational fields, which in turn can lead to the emergence of phenomena that cannot be fully explained within GR alone. Rastall's theory has been widely studied in cosmological contexts and has shown promise in explaining dark energy and certain aspects of black hole physics \cite{29b,30b,31b,37b,38b}. 

The generalized Einstein field equation in Rastall gravity is given by:
\begin{equation}\label{2}
G_{\mu\nu} + \kappa \lambda g_{\mu\nu} R = \kappa T_{\mu\nu},
\end{equation}
where \(\kappa = 8\pi G_N\) is the gravitational constant in Rastall gravity, and \(G_N\) is the gravitational constant in Newtonian gravity. When \(\lambda \rightarrow 0\), this field equation reduces to the Einstein field equation in GR, which reflects the traditional view of energy conservation in the context of gravity.

To further explore the physical implications of this theory, we investigate a charged Reissner–Nordström black hole solution surrounded by perfect fluid matter. In this context, the perfect fluid is characterized by an equation of state \(p = \omega \rho\), where \(p\) is the pressure, \(\rho\) is the density, and \(\omega\) is the equation of state parameter that determines the nature of the fluid. The presence of perfect fluid matter around the black hole is modeled by a specific term in the metric functions, as discussed below.

The metric of a charged black hole surrounded by perfect fluid matter in Rastall gravity is written as \cite{34b}:
\begin{equation}\label{3}
ds^2 = -f(r)dt^2 + g^{-1}(r)dr^2 + r^2(d\theta^2 + \sin^2\theta d\phi^2),
\end{equation}
where the metric functions \(f(r)\) and \(g(r)\) are given by \cite{34b,34c,34d}:
\begin{equation}\label{4}
f(r) = g(r) = 1 - \frac{2M}{r} + \frac{Q^2}{r^2} - \alpha r^{-\frac{1 + 3\omega - 6k\gamma(1 + \omega)}{1 - 3k\gamma(1 + \omega)}},
\end{equation}
Here, \(M\) is the mass of the black hole, \(Q\) is the electric charge, and \(\alpha\) is a constant representing the intensity of the perfect fluid surrounding the black hole. The parameters \(k\) and \(\gamma\) arise from the Rastall gravitational modifications, while \(\omega\) represents the equation of state parameter of the perfect fluid. The gauge potential associated with the electromagnetic field of the charged black hole is:
\begin{equation}
A = \frac{Q}{r} dt,
\end{equation}
which corresponds to a radial electric field. This general form of the solution is consistent with the standard Reissner-Nordstr\"{o}m black hole in the limit where \(\alpha \rightarrow 0\) and \(\lambda \rightarrow 0\), reducing the theory to Einstein’s General Relativity. 

In the presence of the perfect fluid, different values of \(\omega\) lead to different physical interpretations. For example, when \(\omega = -1/3\), the solution describes a black hole surrounded by dark energy, which is known to generate asymptotically flat rotation curves for galaxies \cite{21b,22b,23b}. For \(\omega = 0\), the black hole is surrounded by a dust field, while \(\omega = 1/3\) describes radiation. Meanwhile, quintessence and phantom fields correspond to \(\omega = -2/3\) and \(\omega = -4/3\), respectively \cite{isAl-Badawi:2019tom}. These various choices of \(\omega\) allow us to study the impact of different types of matter surrounding the black hole on its thermodynamics and extremality conditions.

In the limit where \(\lambda \rightarrow 0\), the above solution for the metric reduces to the Kiselev black hole formalism, which describes black holes surrounded by dark energy or quintessence within the framework of GR \cite{49b,50b}. In this formalism, the effects of surrounding perfect fluid matter are encoded in a simple power-law term that modifies the metric functions. Kiselev's solution has been widely applied in the study of black holes with dark energy or quintessence fields.

The interplay between the Rastall parameters \(k\) and \(\gamma\), as well as the perfect fluid parameter \(\alpha\), provides a rich structure for the analysis of black hole extremality. The presence of these additional parameters modifies the usual Reissner-Nordstr\"{o}m black hole solution and leads to significant changes in the black hole's thermodynamics and its extremal limit. By solving the equation \(f(r) = 0\), we can determine the location of the event horizon, and from there, we can study the conditions under which the black hole satisfies the WGC.

In the following sections, we will present a detailed analysis of the photon spheres and their role in validating the WGC in this modified theory of gravity. Additionally, we will study how the presence of perfect fluid matter affects the extremality conditions in black hole structure in Rastall gravity.

\section{Weak Gravity Conjecture}
In this section, we examine the WGC in the context of charged black holes surrounded by perfect fluid in Rastall gravity. The WGC asserts that in any consistent theory of quantum gravity, gravity must be the weakest force, implying that there exists a particle with a charge-to-mass ratio greater than that of extremal black holes. For charged black holes, this means that the charge \(Q\) must exceed the mass \(M\) for extremal black holes to remain unstable, ensuring that they can discharge by emitting particles. The condition \((Q/M)_{ext} > 1\) must be satisfied in extremal black holes, where \(T=0\), to avoid the violation of the WGC \cite{a',b',c',d',e',f',g',h'}.

To study the WGC in this context, we consider the metric of a Reissner-Nordström black hole surrounded by perfect fluid matter within the framework of Rastall gravity. The solution is given by Eq. (\ref{4}) with the metric function:
\begin{equation}\label{44a}
1-\frac{2 M}{r}+\frac{Q^2}{r^2} = \alpha r^{-\frac{-6 \gamma k (\omega+1)+3 \omega+1}{1-3 \gamma  k (\omega+1)}}.
\end{equation}
The left-hand side of this equation corresponds to the standard Reissner-Nordström black hole, while the right-hand side introduces corrections due to the surrounding perfect fluid, characterized by the intensity parameter \(\alpha\), Rastall parameters \(k\) and \(\gamma\), and the equation of state parameter \(\omega\). 

To find the extremal condition, we need to locate the event horizon, which can be determined by solving \(f(r) = 0\). Typically, the horizons of a charged black hole are given by:
\begin{equation}\label{horizons}
r_{\pm} = M \pm \sqrt{M^2 - Q^2},
\end{equation}
where \(r_+\) represents the outer horizon and \(r_-\) the inner horizon. When \(Q^2 > M^2\), the black hole becomes overcharged, and no event horizon forms, leading to a naked singularity. This scenario violates the WCCC, which asserts that singularities must be hidden behind event horizons \cite{f',g',h'}.

By rewriting Eq. (\ref{44a}), we separate the two terms as:
\begin{equation}
f_L =1- \frac{2 M}{r} + \frac{Q^2}{r^2},
\end{equation}
and
\begin{equation}
f_R = \alpha r^{-\frac{-6 \gamma k (\omega+1)+3 \omega+1}{1-3 \gamma  k (\omega+1)}}.
\end{equation}
The extremal state is defined by the condition that these two curves are tangent at a specific radius \(r_0\), which allows us to determine the critical values of the parameters that satisfy the extremal condition and test the WGC.

To find the extremality condition, we first solve for the tangency point \(r_0\) by equating the slopes of the two curves:
\begin{equation}\label{44b}
1-\frac{2 M}{r_0}+\frac{Q^2}{r_0^2} = \alpha r_0^{-\frac{-6 \gamma k (\omega+1)+3 \omega+1}{1-3 \gamma k (\omega+1)}},
\end{equation}
and
\begin{equation}\label{44c}
\frac{2 Q^2}{r_0^3}-\frac{2 M}{r_0^2}=\frac{\alpha (-6 \gamma k (\omega+1)+3 \omega+1) r_0^{-\frac{-6 \gamma k (\omega+1)+3 \omega+1}{1-3 \gamma k (\omega+1)}-1}}{1-3 \gamma k (\omega+1)}.
\end{equation}
By solving these two equations, we can determine the extremal radius \(r_0\) and the corresponding value of the perfect fluid intensity \(\alpha_{ext}\). The expressions for these quantities are given as:
\begin{equation}\label{44d}
r_0 = \frac{\left(3 \gamma k M (\omega+1) \pm \sqrt{9 M^2 (\omega-\gamma k (\omega+1))^2+Q^2 (3 \omega-1) (6 \gamma k (\omega+1)-3 \omega-1)}\right)-3 M \omega}{6 \gamma k (\omega+1)-3 \omega-1},
\end{equation}
and
\begin{equation}\label{44e}
\alpha_{ext} = \left(-2 M r_0 + Q^2 + r_0^2\right) r_0^{\frac{3 \omega}{1-3 \gamma k (\omega+1)}-\frac{6 \gamma k (\omega+1)}{1-3 \gamma k (\omega+1)}+\frac{1}{1-3 \gamma k (\omega+1)}-2}.
\end{equation}

\begin{figure}[h!]
 \begin{center}
 \subfigure[]{
 \includegraphics[height=6cm,width=9cm]{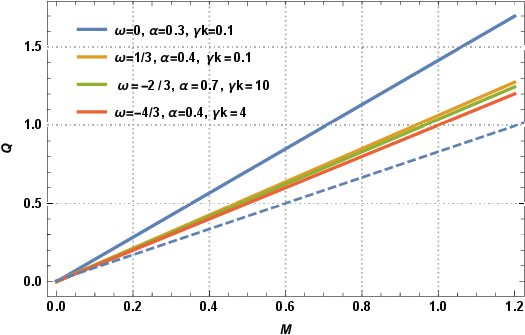}
 \label{21a}}
  \caption{\small{The plot of $(Q/M)$ for the charged black holes in the perfect fluid within Rastall theory for free parameters. The dashed line represents the Q=M}}
 \label{n1}
 \end{center}
 \end{figure}

For extremal black holes, we also compute the charge-to-mass ratio \((Q/M)_{ext}\) in terms of the parameters \(\alpha\), \(\gamma\), and \(k\). By keeping the parameters \(\alpha\), \(k\), \(\gamma\), \(M\), and \(Q\) constant, and increasing \(Q^2/M^2\) until it reaches the critical value \((Q^2/M^2)_{ext}\), we find the following relation:
\begin{equation}\label{44f}
\begin{split}
\left(\frac{M^2}{Q^2}\right)_{ext} =& \bigg[\alpha \bigg( \frac{3 (\gamma k (\omega+1) \pm \omega - \gamma k (\omega+1) - \omega)}{6 \gamma k (\omega+1) - 3\omega - 1} \bigg)^{-\frac{3\omega-1}{3 \gamma k (\omega+1) - 1} + 2}\\&\bigg/\bigg(\frac{\alpha (-6 \gamma k (\omega+1) + 3\omega + 1)^2}{\left(3 M (\gamma k (\omega+1) \pm \omega - \gamma k (\omega+1) - \omega)\right)^2} - 1\bigg) \bigg]
\end{split}.
\end{equation}
Further simplification leads to:
\begin{equation}\label{44g}
\left(\frac{Q^2}{M^2}\right)_{ext} = 1 + \bigg(-\alpha \bigg[3 (-\gamma k (\omega+1) + (\gamma k (\omega+1) \pm \omega) - \omega)\bigg]^{\frac{3 \omega-1}{3 \gamma k (\omega+1) - 1} - 2}\times\bigg[6 \gamma k (\omega+1) - 3 \omega - 1\bigg]^{-1}\bigg).
\end{equation}
So, We assume the expression inside the parentheses is a constant expression \( \sigma \), hence, $
\left(\frac{Q^2}{M^2}\right)_{ext} = 1 + \sigma
$. 
These results allow us to evaluate the validity of the WGC for black holes in the presence of perfect fluid in Rastall gravity. If \((Q/M)_{ext} > 1\), then the WGC holds, indicating that gravity is indeed the weakest force in this scenario. In Fig. (\ref{n1}), the variation of the charge-to-mass ratio \((Q/M)\) is shown as a function of selected free parameters, based on Eq. (\ref{44g}). In the following section, we will explore how the photon spheres surrounding the black holes provide additional evidence for the WGC, particularly in relation to the WCCC.

\section{Photon Spheres and the Weak Gravity Conjecture}
In this section, we aim to investigate the WGC in the context of charged black holes surrounded by perfect fluid in Rastall theory, focusing particularly on the role of photon spheres (PSs). Photon spheres are regions where null geodesics, or photon orbits, can exist, providing important insights into the stability and structure of black holes. In this study, we aim to understand how the extremality conditions of these black holes, characterized by \((Q/M)_{ext} > 1\), validate the WGC and how the presence of PSs reflects the fundamental properties of black holes.

We start by considering the regular potential \(H(r, \theta)\), defined by:
\begin{equation}\label{5}
H(r, \theta) = \sqrt{\frac{-g_{tt}}{g_{\varphi\varphi}}} = \frac{1}{\sin\theta}\left(\frac{f(r)}{h(r)}\right)^{1/2},
\end{equation}
where the metric functions \(f(r)\) and \(h(r)\) describe the spacetime geometry around the black hole. The radius of the PS is located at the point where the radial derivative of \(H(r, \theta)\) vanishes, i.e.,
\begin{equation}\label{6}
\partial_r H = 0.
\end{equation}
To further analyze the PS structure, we introduce the vector field \(\phi = (\phi^r, \phi^\theta)\), with components defined as:
\begin{equation}\label{7}
\phi^r = \frac{\partial_r H}{\sqrt{g_{rr}}} = \sqrt{g(r)}\partial_r H, \quad \phi^\theta = \frac{\partial_\theta H}{\sqrt{g_{\theta\theta}}} = \frac{\partial_\theta H}{\sqrt{h(r)}}.
\end{equation}

We can also define the winding number, which is an important topological property, using the following formula:
\begin{equation}\label{8}
\omega_i = \frac{1}{2\pi}\int_{C_i} d\Lambda,
\end{equation}
where \(\Lambda = \frac{\phi^2}{\phi^1}\). The total topological charge associated with the PSs is then given by the sum of the winding numbers:
\begin{equation}\label{9}
Q = \sum_{i}\omega_i.
\end{equation}

By solving Eqs. (\ref{4}), (\ref{5}), and (\ref{7}), we find the components of the vector field \(\phi^r\) and \(\phi^\theta\):
\begin{equation}\label{10}
\phi^r = \frac{3 \alpha (\omega+1) \csc (\theta ) (4 \gamma k -1) r^{\frac{3 \omega-1}{3 \gamma k (\omega+1)-1}-4}}{2 (3 \gamma k (\omega+1)-1)} - \frac{\csc (\theta ) (r-3 M)}{r^3} - \frac{2 Q^2 \csc (\theta )}{r^4},
\end{equation}
and
\begin{equation}\label{11}
\phi^{\theta } = -\frac{\cot (\theta ) \csc (\theta ) \sqrt{\frac{-\alpha r^{\frac{3 \omega-1}{3 \gamma k (\omega+1)-1}}-2 M r+Q^2+r^2}{r^2}}}{r^2}.
\end{equation}

The presence of a zero point within a closed curve indicates that the total charge \(Q\) is equal to the winding number. As every black hole with a photon sphere likely possesses a topological charge at its PS, we can assign a distinct topological charge to each PS, which can either be \(+1\) or \(-1\). Furthermore, depending on the choice of the closed curve, which may encompass one or more zero points, the total charge may be \(-1\), \(0\), or \(+1\), as shown in \cite{21}.

Based on these ideas, we now investigate the WGC in relation to the PS structure of charged black holes surrounded by perfect fluid within Rastall theory. We aim to test whether the WGC holds, taking into account the WCCC for various parameter values. In addition to providing detailed explanations, the results are also summarized in Table (\ref{P1}).

It is well-known that in the standard case, where \(M > Q\), the black hole should exhibit typical characteristics, with a total charge of PS \(= -1\). Building on this, we examine cases where \(M < Q\), exploring various parameter values. It is essential to verify that, in the \(Q > M\) scenario, the structure satisfies both the WCCC and the PS conditions, ensuring that the black hole retains its structure with an event horizon and a PS having a total negative charge of \(-1\). Based on these free parameter values, we thoroughly analyze the behavior of the structure in the subsequent figures.

In Fig. (\ref{n1}), the variation of the charge-to-mass ratio \((Q/M)\) is shown as a function of selected free parameters, based on Eq. (\ref{44g}).

Next, we study the effect of the PS on the WGC. We plot the behavior of the PS structure for different values of the parameters \(\alpha\), \(k\gamma\), and \(\omega\). The results show how the intensity of the perfect fluid, \(\alpha\), impacts the black hole structure. As \(\alpha\) increases while holding the other parameters constant, we observe significant changes in the system.

For the dust field (\(\omega = 0\)), despite satisfying the WGC condition at extremality (\(T = 0\) with \((Q/M)_{ext} > 1\)), the total charge of the PS remains \(PS = -1\), indicating the presence of a black hole. However, further increases in \(k\) and \(\lambda\) decreases of \(\alpha\) cause the total charge of the PS to become zero, indicating the formation of a singularity instead of a black hole.

\begin{figure}[h!]
 \begin{center}
 \subfigure[]{
 \includegraphics[height=5.5cm,width=5.5cm]{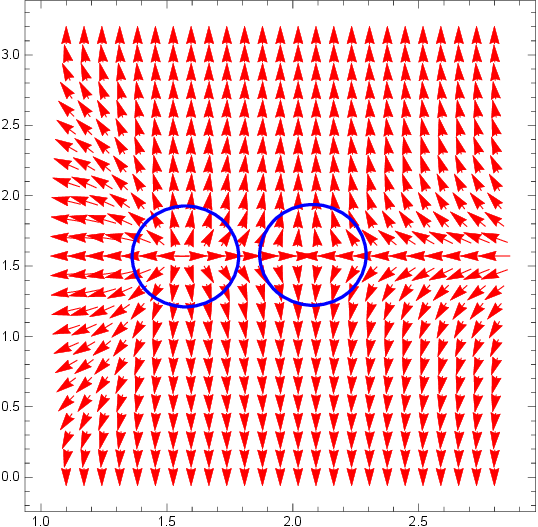}
 \label{2a}}
 \subfigure[]{
 \includegraphics[height=5.5cm,width=5.5cm]{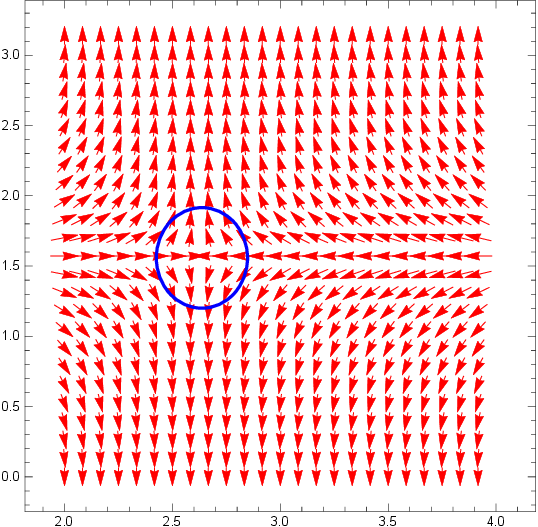}
 \label{2b}}
 \subfigure[]{
 \includegraphics[height=5.5cm,width=5.5cm]{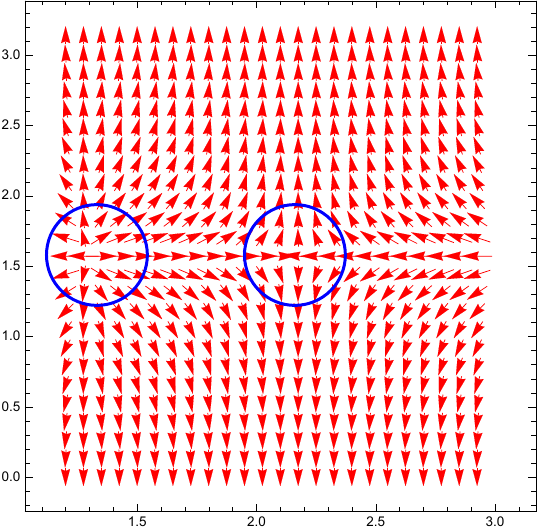}
 \label{2c}}
  \caption{\small{The plot of PSs of charged black holes in the perfect fluid within Rastall theory for \((Q/M)_{ext}>1\) with respect to \((k\gamma=0.1, \alpha=0.1, \omega=1/3)\), \((k\gamma=0.1, \alpha=0.4, \omega=1/3)\), and \((k\gamma=4, \alpha=0.3, \omega=1/3)\), respectively.}}
 \label{m2}
 \end{center}
 \end{figure}

\begin{figure}[h!]
 \begin{center}
 \subfigure[]{
 \includegraphics[height=5.5cm,width=5.5cm]{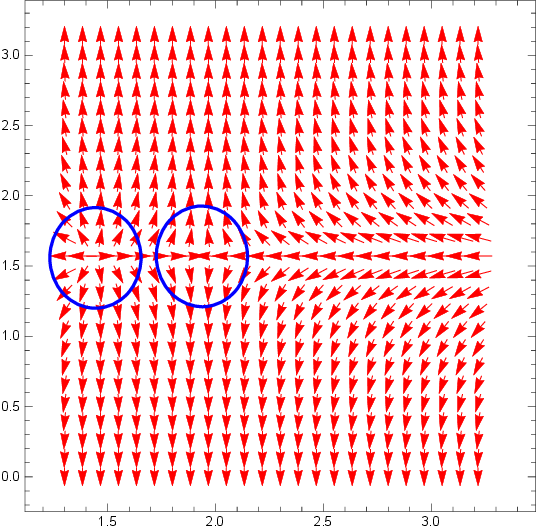}
 \label{3a}}
 \subfigure[]{
 \includegraphics[height=5.5cm,width=5.5cm]{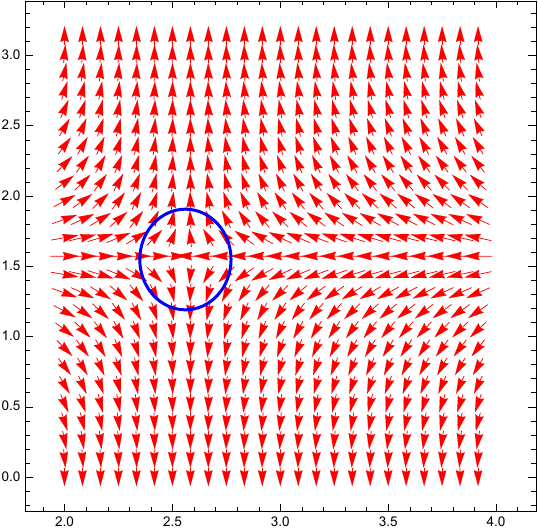}
 \label{3b}}
 \subfigure[]{
 \includegraphics[height=5.5cm,width=5.5cm]{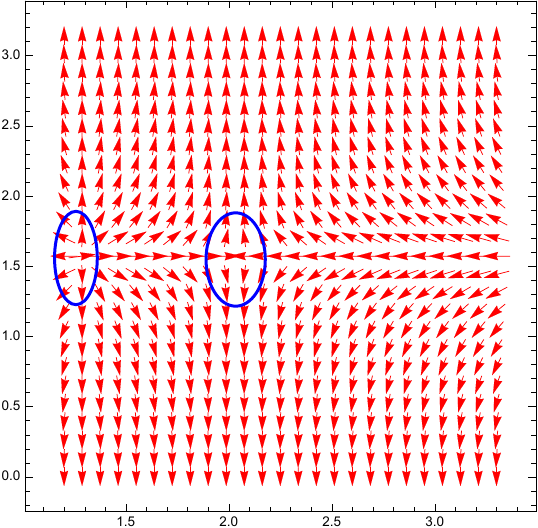}
 \label{3c}}
  \caption{\small{The plot of PSs of charged black holes in the perfect fluid within Rastall theory for \((Q/M)_{ext}>1\) with respect to different values of \((k\gamma=12, \alpha=0.25, \omega=-2/3)\), \((k\gamma=10, \alpha=0.7, \omega=-2/3)\), and \((k\gamma=9, \alpha=0.3, \omega=-2/3)\), respectively.}}
 \label{m3}
 \end{center}
 \end{figure}

\begin{figure}[h!]
 \begin{center}
 \subfigure[]{
 \includegraphics[height=5.5cm,width=5.5cm]{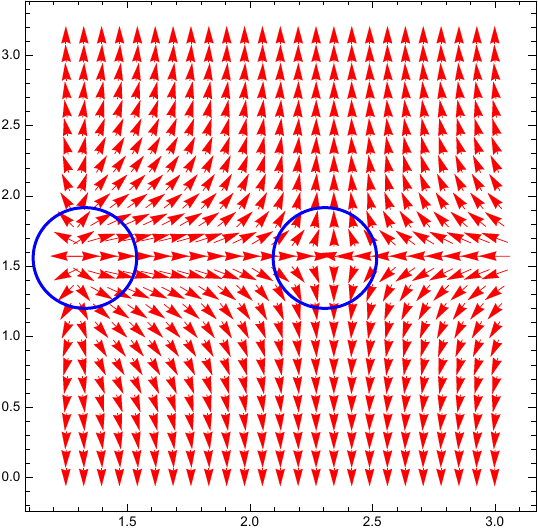}
 \label{4a}}
 \subfigure[]{
 \includegraphics[height=5.5cm,width=5.5cm]{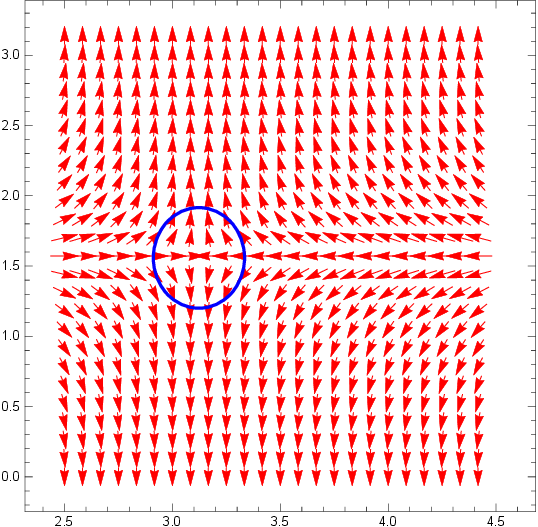}
 \label{4b}}
 \subfigure[]{
 \includegraphics[height=5.5cm,width=5.5cm]{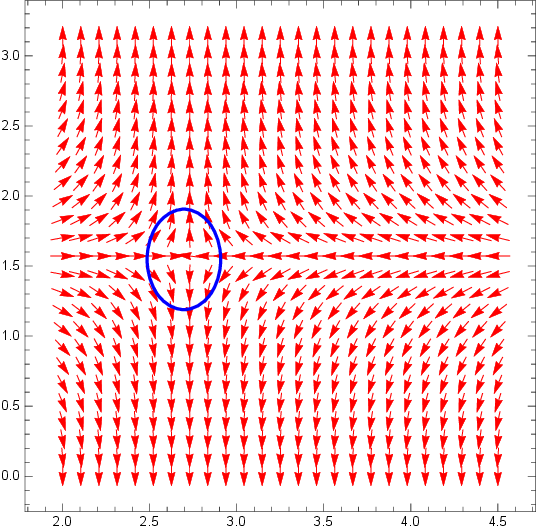}
 \label{4c}}
  \caption{\small{The plot of PSs of charged black holes in the perfect fluid within Rastall theory for \((Q/M)_{ext}>1\) with respect to different values of \((k\gamma=10, \alpha=0.3, \omega=-4/3)\), \((k\gamma=4, \alpha=0.5, \omega=-4/3)\), and \((k\gamma=10, \alpha=0.5, \omega=-4/3)\), respectively.}}
 \label{m4}
 \end{center}
 \end{figure}

Similar behavior is observed in Fig. (\ref{m2}) for the radiation field (\(\omega = 1/3\)). The quintessence field (\(\omega = -2/3\)) and phantom field (\(\omega = -4/3\)) display slightly different behaviors, as shown in Figs. (\ref{m3}) and (\ref{m4}). In these cases, changes in the parameters \(\alpha\), \(k\), and \(\lambda\) effectively confirm the WGC while maintaining a total charge of \(PS = -1\).

The numerical solutions provide evidence for points satisfying the WGC in the extreme limit of the black hole (\(T = 0\)) where \((Q/M)_{ext} > 1\) and \(PS = -1\), indicating the presence of a black hole. The results are summarized in Table (\ref{P1}).

\begin{figure}[h!]
 \begin{center}
 \subfigure[]{
 \includegraphics[height=5.5cm,width=5.5cm]{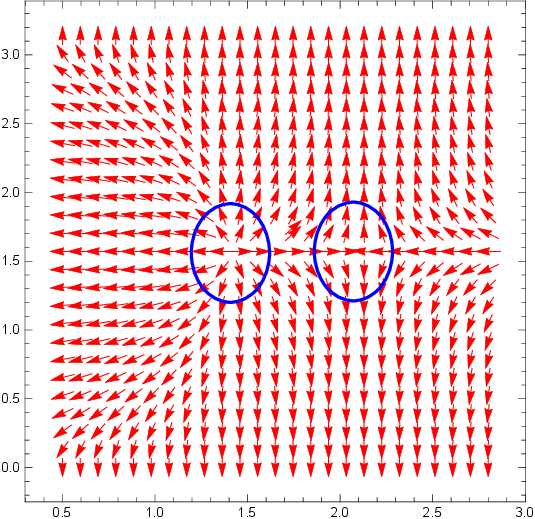}
 \label{1a}}
 \subfigure[]{
 \includegraphics[height=5.5cm,width=5.5cm]{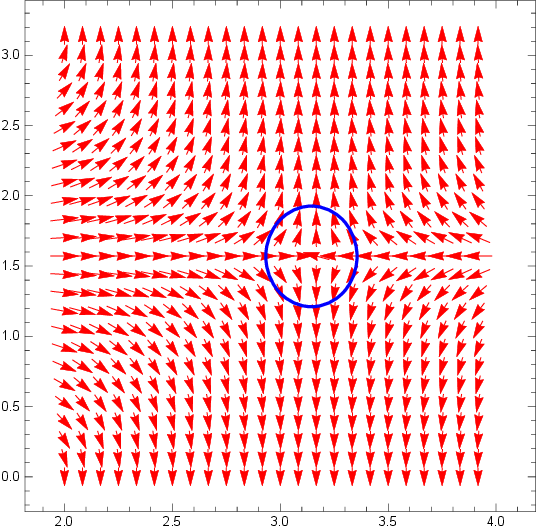}
 \label{1b}}
 \subfigure[]{
 \includegraphics[height=5.5cm,width=5.5cm]{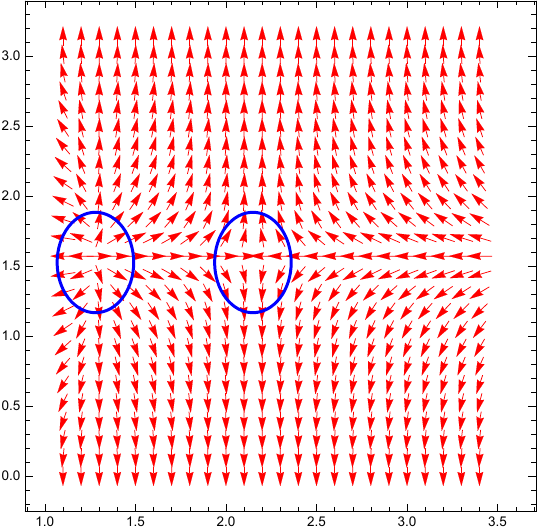}
 \label{1c}}
  \caption{\small{The plot of PSs of charged black holes in the perfect fluid within Rastall theory for \((Q/M)_{ext}>1\) with respect to different values of \((k\gamma=0.1, \alpha=0.1, \omega=0)\), \((k\gamma=0.1, \alpha=0.3, \omega=0)\), and \((k\gamma=10, \alpha=0.3, \omega=0)\), respectively.}}
 \label{m1}
 \end{center}
 \end{figure}

As shown in Fig. (\ref{m1}), as the intensity parameter \(\alpha\) increases while the other parameters remain fixed, significant changes in the system occur. For dust fields (\(\omega = 0\)), the WGC is satisfied with \((Q/M)_{ext}>1\), and the PS total charge is \(-1\), indicating the black hole structure. However, for larger values of \(k\) and \(\lambda\), the total PS charge becomes zero, indicating a singularity rather than a black hole.

The behaviors for the radiation field (\(\omega = 1/3\)) and the quintessence (\(\omega = -2/3\)) and phantom fields (\(\omega = -4/3\)) are displayed in Figs. (\ref{m2}), (\ref{m3}), and (\ref{m4}). These results show that as the parameters \(\alpha\), \(k\), and \(\lambda\) change, the WGC is still satisfied while maintaining a total charge of \(PS = -1\). 

\begin{center}
\begin{table}[H]
  \centering
\begin{tabular}{|p{1cm}||p{1cm}||p{1cm}||p{4cm}||p{2.6cm}||p{2.5cm}||p{3cm}|}
  \hline
   \hspace{0.2cm} $k\gamma$ & \hspace{0.1cm} $\alpha$ & \hspace{0.1cm} $\omega$ & \hspace{0.1cm} PS Total charge  & \hspace{0.6cm} $(Q/M)_{ext}$ & \hspace{0.6cm} WGC & \hspace{0.4cm} PS-WGC\\[3mm]
   \hline
    0.1& 0.1&  0 &\hspace{1.6cm} 0  & \hspace{0.5cm} $1.15531$ & \hspace{1cm} $\surd$ & \hspace{1cm} $\times$\\[3mm]
   \hline
    0.1&  0.3&  0&\hspace{1.5cm}  -1 & \hspace{0.5cm} $1.41571$ & \hspace{1cm} $\surd$ & \hspace{1cm} $\surd$ \\[3mm]
  \hline
    10&  0.3&  0&\hspace{1.6cm}  0  & \hspace{0.5cm} $0.999999$ & \hspace{1cm} $\times$ & \hspace{1cm}  $\times$ \\[3mm]
  \hline
    0.1&  0.1&  1/3&\hspace{1.6cm}  0  & \hspace{0.5cm} $1.01615$ & \hspace{1cm} $\surd$ & \hspace{1cm} $\times$ \\[3mm]
  \hline
   0.1&  0.4&  1/3&\hspace{1.5cm}  -1 & \hspace{0.5cm} $1.06311$ & \hspace{1cm} $\surd$ & \hspace{1cm} $\surd$ \\[3mm]
  \hline
    4&  0.3&  1/3&\hspace{1.6cm}  0 & \hspace{0.5cm} $0.999974$ & \hspace{1cm} $\times$ & \hspace{1cm} $\times$ \\[3mm]
  \hline
   12&  0.25&  -2/3&\hspace{1.6cm}  0  & \hspace{0.5cm} $0.999999$ & \hspace{1cm} $\times$ & \hspace{1cm} $\times$ \\[3mm]
  \hline
    10&  0.7&  -2/3&\hspace{1.5cm}  -1  & \hspace{0.5cm} $1.000098$ & \hspace{1cm} $\surd$ & \hspace{1cm} $\surd$ \\[3mm]
  \hline
    9&  0.3&  -2/3&\hspace{1.6cm} 0  & \hspace{0.5cm} $0.99999984$ & \hspace{1cm} $\times$ & \hspace{1cm} $\times$ \\[3mm]
  \hline
    10&  0.3&  -4/3&\hspace{1.6cm}  0  & \hspace{0.5cm} $1.0000039$ & \hspace{1cm} $\surd$ & \hspace{1cm} $\times$ \\[3mm]
  \hline
    4&  0.5&  -4/3&\hspace{1.5cm}  -1  & \hspace{0.5cm} $1.00078$ & \hspace{1cm} $\surd$ & \hspace{1cm} $\surd$ \\[3mm]
  \hline
   10&  0.5&  -4/3&\hspace{1.5cm}  -1  & \hspace{0.5cm} $1.00001$ & \hspace{1cm} $\surd$ & \hspace{1cm} $\surd$ \\[3mm]
  \hline
\end{tabular}
\caption{Summary of the results of the WGC validation with PS monitoring}\label{P1}
\end{table}
\end{center}

Therefore, the presence of a photon sphere with total charge PS\( = -1\) serves as an indicator that the black hole satisfies the WGC, as demonstrated in our analysis of different values for the free parameters. However, in certain regions of the parameter space, the photon sphere disappears, leading to the breakdown of the black hole structure and the formation of naked singularities, violating the WCCC. These results provide further evidence for the WGC and offer valuable insights into the interaction between quantum gravity and the cosmic structure of black holes in modified theories such as Rastall gravity.

\section{Thermodynamic Topology}
Recently, innovative methods have emerged to analyze and compute critical points and phase transitions in black hole thermodynamics. One notable approach is the topological method. To adopt a topological perspective in thermodynamics, Duan’s topological current \(\phi\)-mapping theory is highly recommended. Wei et al. have introduced distinct methods for studying topological thermodynamics, focusing on generalized free energy functions. This approach treats black holes as defects in the thermodynamic parameter space, with their solutions explored using the generalized off-shell free energy. In this framework, the stability and instability of black hole solutions are indicated by positive and negative winding numbers, respectively.

To explore the thermodynamic properties of black holes, various quantities are employed. For instance, mass and temperature can describe the generalized free energy. Given the relationship between mass and energy in black holes, we express the generalized free energy function as a standard thermodynamic function in the following form \cite{18a,19a}:
\begin{equation}\label{12}
\mathcal{F} = M - \frac{S}{\tau},
\end{equation}
where \(\tau\) denotes the Euclidean time period, while \(T\) (the inverse of \(\tau\)) represents the temperature of the ensemble. The generalized free energy is on-shell only when \(\tau = \tau_{H} = \frac{1}{T_{H}}\). As stated in \cite{18a,19a}, a vector \(\phi\) is constructed as follows:
\begin{equation}\label{13}
\phi = \left(\frac{\partial\mathcal{F}}{\partial r_{H}}, -\cot\Theta\csc\Theta\right).
\end{equation}
Here, \(\phi^{\Theta}\) diverges, and the vector direction points outward at \(\Theta = 0\) and \(\Theta = \pi\). The ranges for \(r_{H}\) and \(\Theta\) are \(0 \leq r_{H} \leq \infty\) and \(0 \leq \Theta \leq \pi\), respectively. Using Duan's \(\phi\)-mapping topological current theory, a topological current can be defined as follows:
\begin{equation}\label{14}
j^{\mu} = \frac{1}{2\pi} \varepsilon^{\mu\nu\rho} \varepsilon_{ab} \partial_{\nu}n^{a} \partial_{\rho}n^{b}, \hspace{1cm} \mu, \nu, \rho = 0, 1, 2,
\end{equation}
where \(n = (n^1, n^2)\), and \(n^1 = \frac{\phi^r}{\|\phi\|}\), \(n^2 = \frac{\phi^\Theta}{\|\phi\|}\). Noether's theorem ensures that the resulting topological currents are conserved:
\begin{equation}\label{15}
\partial_{\mu}j^{\mu} = 0.
\end{equation}
To determine the topological number, we reformulate the topological current as follows \cite{18a,19a}:
\begin{equation}\label{16}
j^{\mu} = \delta^{2}(\phi) J^{\mu}\left(\frac{\phi}{x}\right),
\end{equation}
where the Jacobi tensor is determined as:
\begin{equation}\label{17}
\varepsilon^{ab}J^{\mu}\left(\frac{\phi}{x}\right) = \varepsilon^{\mu\nu\rho} \partial_{\nu}\phi^{a} \partial_{\rho}\phi^{b}.
\end{equation}
The Jacobi vector reduces to the standard Jacobi when \(\mu = 0\), as demonstrated by \(J^{0}\left(\frac{\phi}{x}\right) = \frac{\partial(\phi^1, \phi^2)}{\partial(x^1, x^2)}\). From Eq. (\ref{15}), we observe that \(j^{\mu}\) is non-zero only when \(\phi = 0\). After some calculations, we can express the topological number or total charge \(W\) as follows:
\begin{equation}\label{18}
W = \int_{\Sigma}j^{0}d^2 x = \sum_{i=1}^{n}\beta_{i}\eta_{i} = \sum_{i=1}^{n}\widetilde{\omega}_{i}.
\end{equation}
Here, \(\beta_i\) denotes the positive Hopf index, which counts the loops of the vector \(\phi^a\) in the \(\phi\)-space when \(x^\mu\) is near the zero point \(z_i\). Meanwhile, \(\eta_i = \text{sign}(j^0(\phi/x)_{z_i}) = \pm 1\). The quantity \(\widetilde{\omega}_i\) represents the winding number for the \(i\)-th zero point of \(\phi\) in \(\Sigma\). Note that the winding number is independent of the shape of the region where the calculation occurs. The value of the winding number is directly related to black hole stability, with a positive (negative) winding number corresponding to a stable (unstable) black hole state.

Based on the above discussion and considering Eqs. (\ref{4}), (\ref{12}), and (\ref{13}), the form of the function \((\phi^{r}, \phi^\Theta)\) is determined as follows:
\begin{equation}\label{19}
\begin{split}
\phi^{r_h} = \frac{1}{2} - \bigg( \frac{3 a (w-\gamma k (w+1)) r^{\frac{3 w-1}{3 \gamma k (w+1)-1}}}{3 \gamma k (w+1)-1} + Q^2 + \frac{4 \pi r^3}{\tau} \bigg) \bigg/ 2 r^2,
\end{split}
\end{equation}
and
\begin{equation}\label{20}
\phi^{\theta } = -\frac{\cot (\theta )}{\sin (\theta )}.
\end{equation}

The unit vectors \( \mathbf{n}_1 \) and \( \mathbf{n}_2 \) are computed using Eq. (\ref{19}). Next, we find the zero points of the \( \phi^{r} \) component by solving \( \phi^{r} = 0 \) and derive an expression for \( \tau \) as follows:
\begin{equation}\label{21}
\begin{split}
\tau = 4 \pi r^3 \times \bigg[ \frac{(3 a \gamma k (w+1) - 3 a w) r^{\frac{3 w-1}{3 \gamma k (w+1)-1}}}{3 \gamma k (w+1)-1} - Q^2 + r^2 \bigg]^{-1}.
\end{split}
\end{equation}

\begin{figure}[h!]
 \begin{center}
 \subfigure[]{
 \includegraphics[height=5.5cm,width=8.5cm]{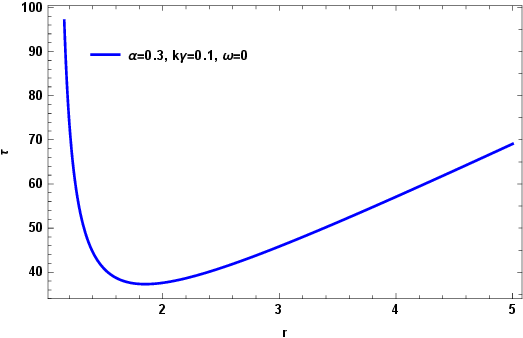}
 \label{5a}}
 \subfigure[]{
 \includegraphics[height=5.5cm,width=8.5cm]{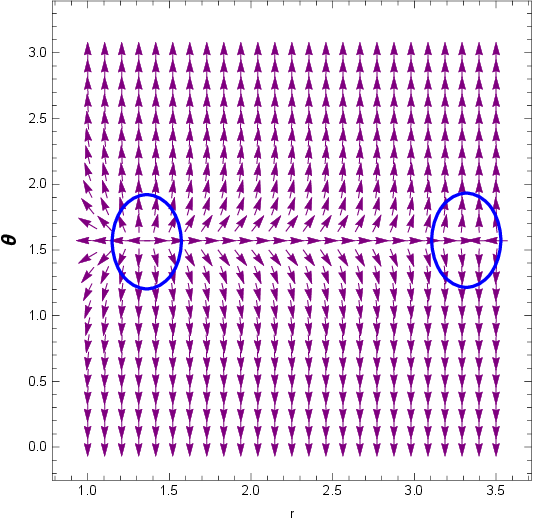}
 \label{5b}}
  \caption{\small{The curve corresponding to Eq.(\ref{21}) is shown in Fig. \ref{5a}. In Fig. \ref{5b}, the zero points (ZPs) are positioned at \((r, \theta)\) on the circular loops with parameters \((\alpha=0.1, k\gamma=0.1, \omega=0)\).}}
 \label{m5}
 \end{center}
 \end{figure}

 \begin{figure}[H]
 \begin{center}
 \subfigure[]{
 \includegraphics[height=5cm,width=8.5cm]{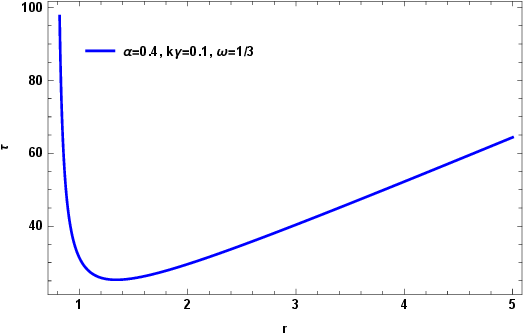}
 \label{6a}}
 \subfigure[]{
 \includegraphics[height=5cm,width=8.5cm]{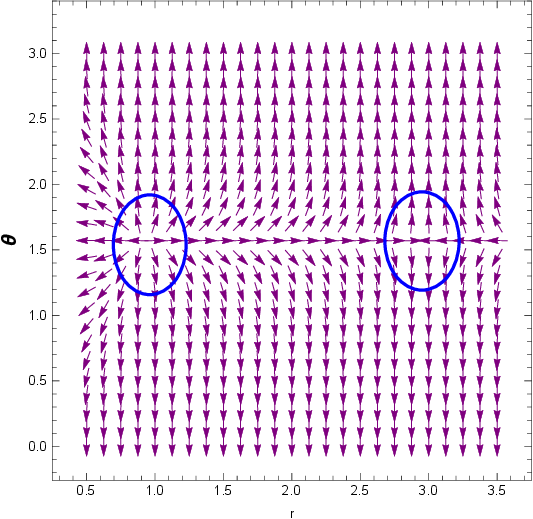}
 \label{6b}}
  \caption{\small{The curve corresponding to Eq.(\ref{21}) is shown in Fig. \ref{6a}. In Fig. \ref{6b}, the ZPs are positioned at \((r, \theta)\) on the circular loops with parameters \((\alpha=0.4, k\gamma=0.1, \omega=1/3)\).}}
 \label{m6}
 \end{center}
 \end{figure}

\begin{figure}[H]
 \begin{center}
 \subfigure[]{
 \includegraphics[height=5cm,width=8.5cm]{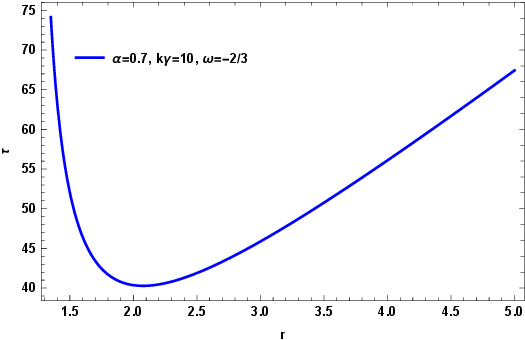}
 \label{7a}}
 \subfigure[]{
 \includegraphics[height=5cm,width=8.5cm]{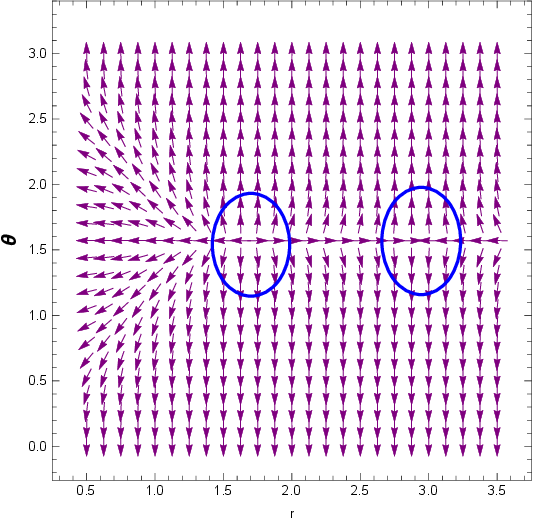}
 \label{7b}}
  \caption{\small{The curve corresponding to Eq.(\ref{21}) is shown in Fig. \ref{7a}. In Fig. \ref{7b}, the ZPs are positioned at \((r, \theta)\) on the circular loops with parameters \((\alpha=0.7, k\gamma=10, \omega=-2/3)\).}}
 \label{m7}
 \end{center}
 \end{figure}

\begin{figure}[h!]
 \begin{center}
 \subfigure[]{
 \includegraphics[height=5cm,width=8.5cm]{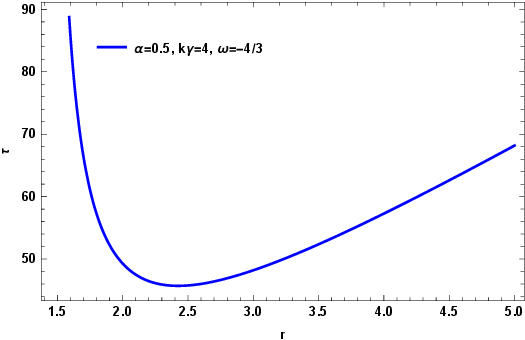}
 \label{8a}}
 \subfigure[]{
 \includegraphics[height=5cm,width=8.5cm]{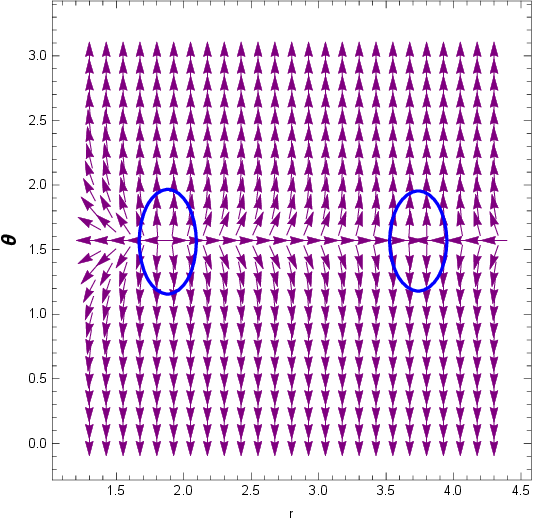}
 \label{8b}}
 \subfigure[]{
 \includegraphics[height=5cm,width=8.5cm]{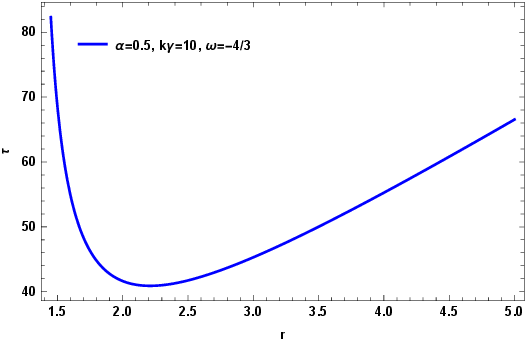}
 \label{8c}}
 \subfigure[]{
 \includegraphics[height=5cm,width=8.5cm]{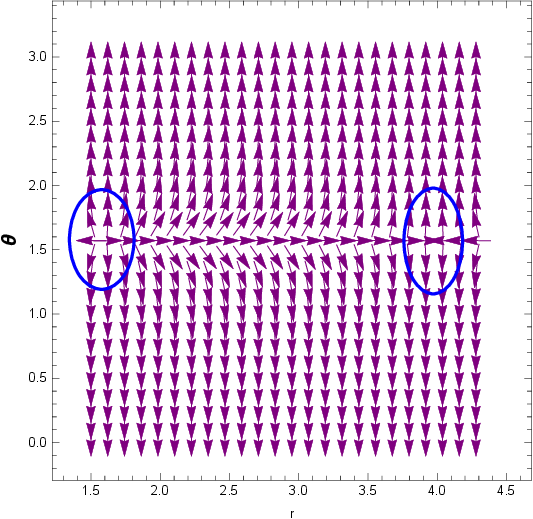}
 \label{8d}}
  \caption{\small{The curve corresponding to Eq.(\ref{21}) is shown in Figs. \ref{8a} and \ref{8c}. In Figs. \ref{8b} and \ref{8d}, the ZPs are positioned at \((r, \theta)\) on the circular loops with parameters \((\alpha=0.5, k\gamma=4, \omega=-4/3)\) and \((\alpha=0.5, k\gamma=10, \omega=-4/3)\), respectively.}}
 \label{m8}
 \end{center}
 \end{figure}

According to the previous section, we identified points where, despite the validation of the WGC, the black hole structure persists due to the existence of photon spheres with a total topological charge of \(-1\). Therefore, in some regions of the free parameters where the aforementioned conditions are met, we will examine the thermodynamic topology of the black hole. We encounter two zero points in Figs. \ref{5b}, \ref{6b}, \ref{7b}, \ref{8b}, and \ref{8d}, indicating two topological charges determined by the free parameters mentioned in the study. These charges correspond to the winding numbers and are located within the blue contour loops at coordinates \((r, \theta)\). The sequence of illustrations is determined by the parameters \(\omega\), \(\gamma\), and \(\alpha\). The findings from all figures reveal the distinctive feature of a total topological charge of \(W=0\) within the contour (In all figures, we encounter topological charges \(\widetilde{\omega} = +1, -1\)).

Generally, for all values of \(\omega\), such as \(0\), \(1/3\), \(-2/3\), and \(-4/3\), changing the free parameters does not alter the number of topological charges. In Figs. \ref{5a}, \ref{6a}, \ref{7a}, \ref{8a}, and \ref{8c}, we plotted the trajectory corresponding to Eq. (\ref{21}) across various free parameter values.

\newpage
\section{Conclusion}
In this paper, we have conducted an in-depth analysis of the WGC within the framework of PSs for charged black holes, considering a perfect fluid environment within the context of Rastall Theory. Our primary objective was to validate the WGC by identifying extremality conditions in these black holes, while also analyzing the role of PSs in establishing a connection between quantum mechanics and general relativity. The intricate interplay between quantum dynamics and gravitational forces observed throughout the study opens up new avenues for exploration in high-energy physics and quantum gravity.

Through the numerical analysis, we have revealed that varying the intensity of the perfect fluid parameter (\(\alpha\)) and the Rastall geometric parameters (\(k\) and \(\lambda\)) leads to significant changes in the black hole structure. Specifically, for dust fields (\(\omega = 0\)), we have observed that the WGC is satisfied under extremality conditions (\(T=0\)) with \((Q/M)_{ext}>1\), indicating a black hole structure as evidenced by the total PS charge of \(-1\). However, as the values of \(k\) and \(\lambda\) increase, the total PS charge becomes zero, representing a singularity rather than a black hole. These findings suggest that while the WGC can be fulfilled, further increases in these parameters cause the black hole structure to collapse into a singularity, making those specific points unfit for the conjecture's validation.

We have also explored other types of perfect fluids, such as the radiation field (\(\omega = 1/3\)), quintessence field (\(\omega = -2/3\)), and phantom field (\(\omega = -4/3\)). Our results confirm that these black hole configurations, framed by the Rastall Theory, satisfy the WGC while maintaining a total PS charge of \(-1\) in specific regions of the free parameters. This ensures that the black hole structure, characterized by a horizon and photon sphere, remains intact even when considering different types of matter distributions.

The novel aspect of this work lies in the dual motivation to investigate both the WGC and the topological structures associated with PSs. We introduced Duan’s topological current \(\phi\)-mapping theory to examine the thermodynamic topology of these black holes. This approach has allowed us to further validate the WGC by analyzing the topological nature of the system in relation to its free parameters. Notably, the generalized Helmholtz free energy method was employed to categorize the topological classes of black holes. Our analysis has identified regions where the total topological charge was \(W=0\).

One of the critical insights gained from this study is the inherent stability of charged black holes surrounded by different types of perfect fluids in Rastall gravity. The existence of stable PSs across different values of the free parameters ensures that the system remains physically viable and adheres to both the WGC and the WCCC. The connection established between quantum gravity and relativistic gravitational systems, as demonstrated through the PS analysis, suggests a new avenue for reconciling quantum mechanics with general relativity in higher-energy regimes.

Given the results obtained in this manuscript, there are several promising directions for future research. One important area would be to explore how higher-dimensional black hole solutions in the context of Rastall gravity interact with the WGC and PSs. Such an extension could provide deeper insights into how extra dimensions influence the charge-to-mass ratio and the formation of PSs, which could have implications for string theory and the landscape of viable low-energy effective theories.

Additionally, we plan to investigate the role of higher-order curvature corrections, such as those arising from Gauss-Bonnet or Lovelock gravity, to understand how these corrections affect the WGC in more complex gravitational backgrounds. It would also be of interest to study black hole solutions in other modified gravity frameworks, like \(f(R)\) gravity or Einstein-Cartan theory, to examine how various forms of modified gravity theories impact the conjectures being tested.

Finally, it is essential to further explore the observational consequences of these theoretical models. By examining gravitational lensing effects, black hole shadows, and PSs in various black hole configurations, we may gain empirical data that either supports or constrains the theoretical predictions made by the WGC and the Rastall theory-based solutions. Future work could also involve applying the methods developed here to explore other conjectures within the Swampland Program, thereby providing a broader understanding of how black holes fit into the landscape of quantum gravity theories.

\section*{Acknowledgements}
\.{I}.S. gratefully acknowledges the networking support from the COST Action CA21106 - COSMIC WISPers in the Dark Universe: Theory, Astrophysics, and Experiments (CosmicWISPers). In addition, \.{I}.S. extends appreciation for the networking support provided by the COST Action CA18108 - Quantum Gravity Phenomenology in the Multi-Messenger Approach (QG-MM), and the COST Action CA22113 - Fundamental Challenges in Theoretical Physics (THEORY-CHALLENGES). Furthermore, \.{I}.S. expresses his sincere thanks to T\"{U}B\.{I}TAK, SCOAP3, and ANKOS for their continued support.

\end{document}